# Impact of national lockdown on COVID-19 deaths in select European countries and the US using a Changes-in-Changes model

Mudit Kapoor (ISI, Delhi) and Shamika Ravi

**Abstract**

In this paper, we estimate the impact of national lockdown on COVID-19 related total and daily deaths, per million people, in select European countries. In particular, we compare countries that imposed a nationwide lockdown (Treatment group); Belgium, Denmark, France, Germany, Italy, Norway, Spain, United Kingdom (UK), and the US, to Sweden (Control group) that did not impose national lockdown using a changes-in-changes (CIC) estimation model. The key advantage of the CIC model as compared to the standard difference-in-difference model is that CIC allows for mean and variance of the outcomes to change over time in the absence of any policy intervention, and CIC accounts for endogeneity in the choice of policy intervention. Our results indicate that in contrast to Sweden, which did not impose a national lockdown, Germany and, to some extent, the US were the two countries where nationwide lockdown had a significant impact on the reduction in COVID-19 related total and daily deaths per million people. In Norway and Denmark, there was no significant impact on total and daily deaths per million people relative to Sweden. Whereas in other countries; Belgium, France, Italy, Spain, and the UK, the effect of the lockdown was in the opposite direction, that is, they experienced significantly higher COVID-19 related total and daily deaths per million people, post the lockdown as compared to Sweden. Our results suggest that the impact of nationwide lockdown on COVID-19 related total and daily deaths per million people varied from one country to another.

**Introduction**

COVID-19 pandemic has infected more than 7 million people and taken more than four hundred thousand lives across the world. Even though the epidemic started in Wuhan, China, it has disproportionately affected Western European and the North American region; together, they account for approximately 50% and 70% of the cases, and deaths, respectively, while only accounting for about 11% of the population. The most common policy response to the pandemic of the countries in regions of Northern America and the Western European region has been to impose a national lockdown of varying measures, which included restrictions on mobility, quarantines, and some form of curfews. In contrast, Sweden made a conscious decision not to impose a national lockdown. Instead, Swedish authorities believed that voluntary self-preventive measures taken by the general public would be sufficient to contain the pandemic. However, in light of the recent excess deaths particularly among the elderly, Swedish response to the epidemic has come under scrutiny.[1,2]

In this paper, we use statistical techniques such as the changes-in-changes (CIC) model developed by Athey and Imbens (2006)[3] to estimate the impact of national lockdown in select Western European and North American countries; Belgium, Denmark, France, Germany, Italy, Norway, Spain, and United Kingdom, and the United States (US), in terms of COVID-19 related total and daily deaths, per million people, by comparing to Sweden that did not impose a national lockdown.

**Data and Methods**

We use data from COVID-19 Dashboard by the Center for Systems Science and Engineering (CSSE) at Johns Hopkins University, referred here as JHU data.[4] JHU data provides daily country-level data on confirmed cases (deaths) which includes presumptive and probable cases (deaths) as per the Center for Disease guidelines on April 14[th].[5] The countries in our sample, include Belgium, Denmark, France, Germany, Italy, Norway, Spain, and the United Kingdom, the United States (US), and Sweden. The time frame for each country was the date when the first case was reported till May 20[th], 2020 (which is the date by which countries had lifted the lockdown).

For each country, we compute COVID-19 related total and daily deaths per million people on a daily basis. The population data for each country is from the open database of the World Bank.[6] We perform a pair-wise comparison for each country that imposed the national lockdown (Treatment group) with Sweden (Control group) that did not impose the nationwide lockdown. We use two dependent variables, total deaths per million people, and daily deaths per million people, to study the impact of the lockdown.

We use *changes-in changes* (CIC) model developed by Athey and Imbens (2006)[3] to estimate the impact of lockdown for each of the countries. The CIC model generalizes the more commonly used differences-in differences (DID) model, which is typically used to measure the impact of policy. In a typical DID model, there is a treatment group, which in our analysis is the country that imposed the national lockdown, and a control group, Sweden, that did not impose the nationwide lockdown. For each of the treatment countries, dates prior to the lockdown are the control period, and dates post the lockdown is the treatment period. There are three distinct advantages of using the CIC model over the DID model; first, the CIC model is a generalization of the DID model. As Athey and Imbens (2006)[3] note that "it allows for the distribution of the unobservable to vary across groups in an arbitrary way." Second, it allows for the possibility that outcomes could change over time without any policy intervention. In our setting, this is important as individual behavior could also change in response to the pandemic (prevalence elasticity of demand for self-protection), as was observed in previous epidemics such as the H1N1 flu (see Bhattacharya, Hyde, and Tu (2013))[7]. Third, the policy intervention is not exogenous; it is more likely to be adopted when the policymakers believe that it will have the most significant impact, the CIC model allows for this possibility while DID assumes exogeneity of policy intervention.

As explanatory variables, we allow for time trend with a structural break. To estimate structural break in the trend for each country, we use the logarithmic value of total deaths per million people and regress it on time for each country and determine the date when there is a structural break in trend.

The CIC model is computed using the *cic* command developed by Keith Kranker, which is Stata implementation of the Athey and Imbens (2006)[3] Changes-in-Changes model in STATA 14.2 (StataCorp LLC, Texas, USA). The standard errors are computed using the bootstrap estimation based on 50 replications. We report mean and median estimates of the CIC with standard errors, results of the DID estimates are also reported. The structural break in the trend is estimated using the STATA command *estat sbsingle*.

**Results**

First, we report the evolution of COVID-19 related total deaths per million people, for each of the countries in the treatment group while comparing it to Sweden (see **figure 1**). We find that Belgium, France, Italy, Spain, and the United Kingdom had higher total deaths per million people as compared to Sweden for all periods post the lockdown. While in Denmark, Germany, Norway, and the US, it was lower compared to Sweden. Next, we report COVID-19 related daily deaths per million people, over time, and for each country compute the 7-day centered moving average, which is highlighted as the red line. We observed that for each of the countries (treatment and control), daily deaths per million people increased reached a peak, and then declined. However, it is interesting to note that for Sweden, the daily deaths per million seems noisier as compared to other countries. Moreover, for Demark, Germany, and Norway, daily deaths per million people remained less than 5 (see **figure 2**).

Our next set of results report the CIC (mean and median) and DID estimates for total deaths per million people, along with the bootstrap standard errors for each of the treatment countries (see **figure 3**). We find that for Germany and the US, the CIC and the DID estimates are negative, and the 95% confidence intervals do not intersect 0, suggesting that there was a significant negative impact on total deaths per million people. However, the effect was higher for Germany as compared to the US. For Demark and Norway, only the DID impact is negative and significant, but the CIC estimates are positive and insignificant for Denmark, and negative and insignificant for Norway. While for other countries, CIC and DID estimates are positive and significant. We also report results, for daily deaths per million people, this is a more noisy variable, which is reflected in wider standard errors (see **figure 4**). However, CIC estimates suggest that there was a significant negative impact only for Germany. While for other countries in terms of the CIC median impact, we did not find any significant effect of the national lockdown as compared to Sweden.

**Discussion**

Using the CIC model, we do a pair-wise comparison of countries that imposed the national lockdown (treatment group) in select Western European countries and the US with Sweden (control group) that did not impose a lockdown. Our results indicate that in terms of COVID-19 related total deaths per million people, the lockdown was effective only in Germany and the US. Based on these estimates of CIC mean, Germany had 127 fewer COVID-19 related deaths per million people. This implies that in Germany, approximately 10529 (129% of 8144) extra lives were saved due to the lockdown as on May 20th, 2020. Similarly, the US would have had 69 fewer COVID-19 related deaths per million people. This implies that in the US, approximately 22541 (24% of 93439) extra lives were saved due to the lockdown as on May 20th, 2020.

If Sweden had imposed the lockdown with a similar effect as Germany, they would have saved on an average of 127 deaths per million people, implying that Sweden would have had 1292 fewer deaths (27% of 4795 deaths) as on May 20th, 2020. However, when we contrast Sweden to countries such as Belgium, France, Italy, Spain, and the UK, then Sweden has done much better in comparison. However, as compared to Demark and Norway, there seems to be no differential impact in Sweden. When we compare Sweden to the US, Sweden would have had 763 fewer deaths (16% of 4795 deaths).

This implies that the impact of national lockdown was not uniform across all countries. Germany and the US seem to have benefitted from the lockdown as compared to Sweden, while Denmark and Norway had no differential impact. Still, in other countries, Belgium, France, Italy, Spain, and the United Kingdom, the effect of lockdown has been in the opposite direction as compared to Sweden. This would perhaps imply that these nations were delayed in their national lockdown decisions, costing too many lives compared to Sweden.

It is essential to put our results in context. CIC models allow for individual behavior to change in the absence of any policy intervention. In other words, these results suggest that elasticity of demand for self-protection is positive. Perhaps, individual demand for self-protection, which responds not only to prevalence but also to the mortality rates, could have played an important role in Sweden to keep the pandemic in check. Even though Sweden could have done much better when compared to Germany and the US, but in contrast to other select Western European countries, it has not performed poorly in terms of COVID-19 related total deaths per million people. Our results suggest that we need more studies using individual-level data to estimate the elasticity of demand for self-protection, which could also play an important role in controlling the epidemic. Results from studies on previous outbreaks in the US; AIDS epidemic, measles, and the seasonal influenza epidemic suggests that elasticity of demand for self-protection is positive.[7] We need similar studies for the COVID-19 pandemic.

**Limitation**

Our paper has several limitations. However, the noteworthy ones are related to data. We do not have individual-level data that would have allowed us to do a more rigorous application of the CIC model. In essence, we are treating all individuals in a country to be homogenous. Several examples in the popular press have revealed a disproportionate impact of the pandemic on race, ethnic minority, gender, age, comorbidities, etc. Further individual level and more rigorous studies would be needed to account for that. Second, there is no consensus on the definition of a COVID-19 death even in Western European countries, and the US (see Mark Handley's discussion on deaths).[8]

**Conclusion**

In this paper, we find that the impact of lockdown was not uniform across all countries. Germany and the US, seemed to have benefitted due to the lockdown as compared to Sweden that did not impose a national lockdown. Though Denmark and Norway have lower COVID-19 related deaths per million people as compared to Sweden, our analysis suggests that this might not be driven by a policy of the national lockdown. However, in other countries such as Belgium, Italy, France, Spain, and the United Kingdom, even though they imposed a nationwide lockdown, it is not clear whether it had the desired impact when compared to Sweden. However, we cannot answer the question on what would have happened had these countries imposed a lockdown sooner as suggested by some experts.[9]

**Figure 1:** **Total deaths per million people over time, Across countries till May 20th, 2020**

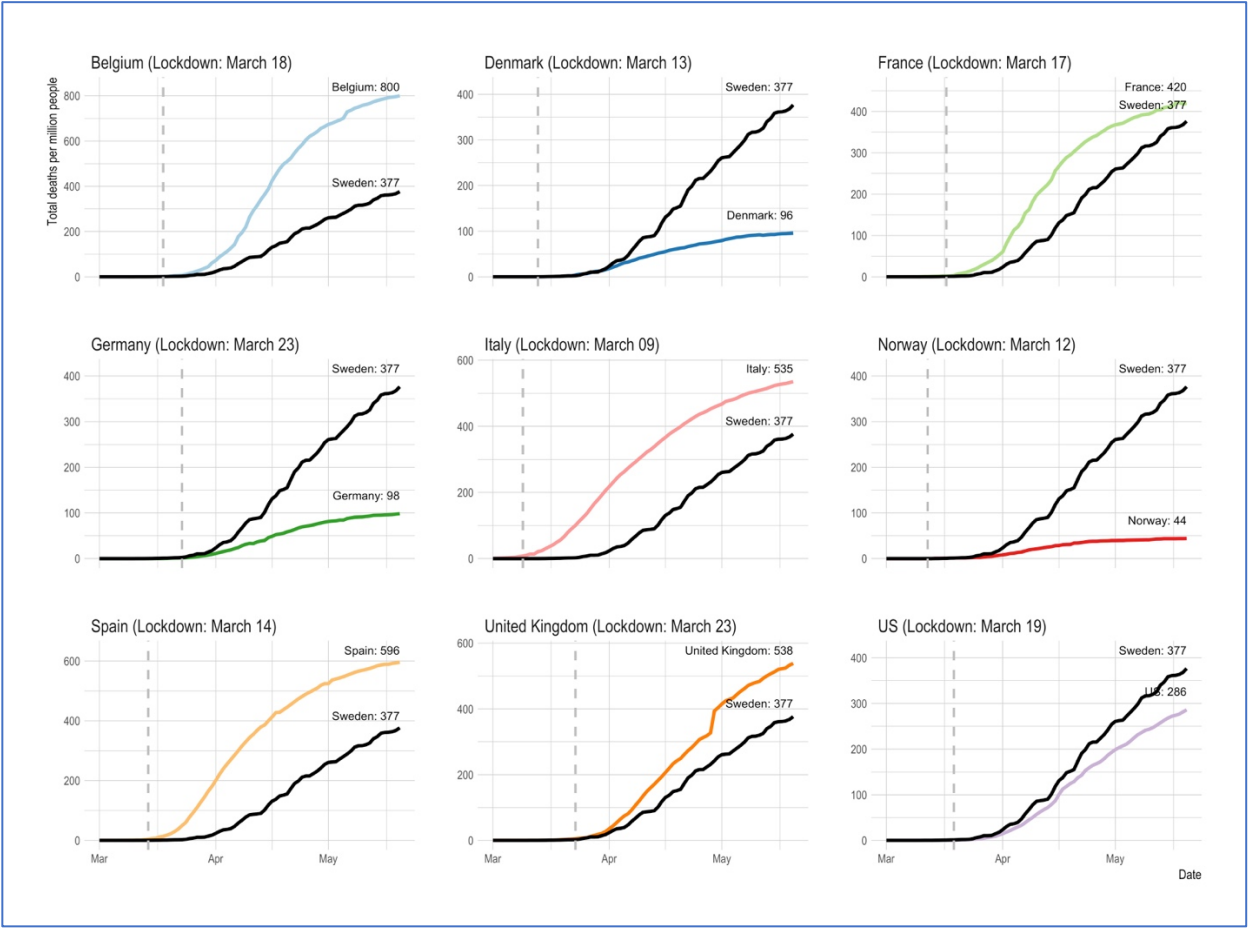

**Figure 2:** **Daily deaths per million people over time, Across countries till May 20th, 2020**

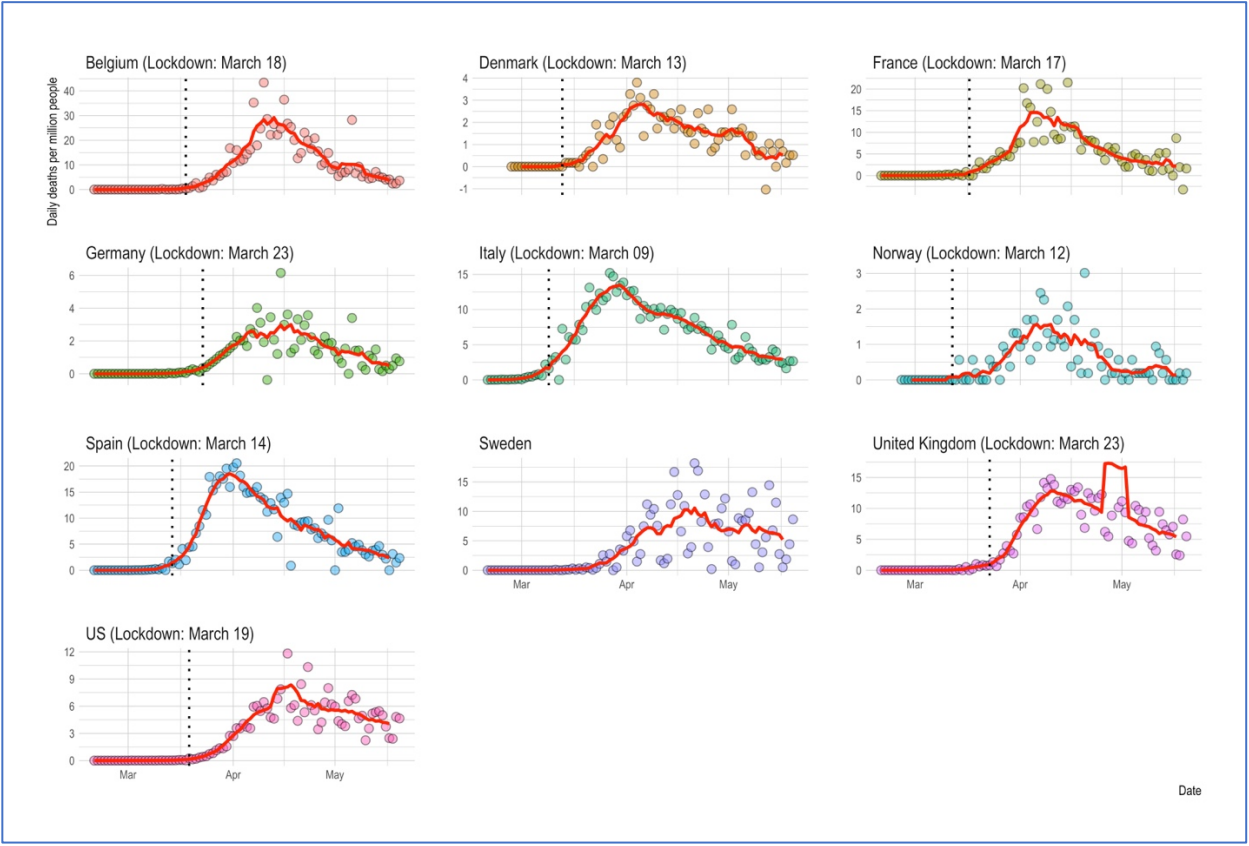

Note: The red line is the 7-day centered moving average. The black dotted line is the start of the national lockdown.

**Figure 3:** Changes-in-Changes (CIC) and Differences-in-Differences (DID) estimates for total deaths per million people, for each country with Sweden as the control group

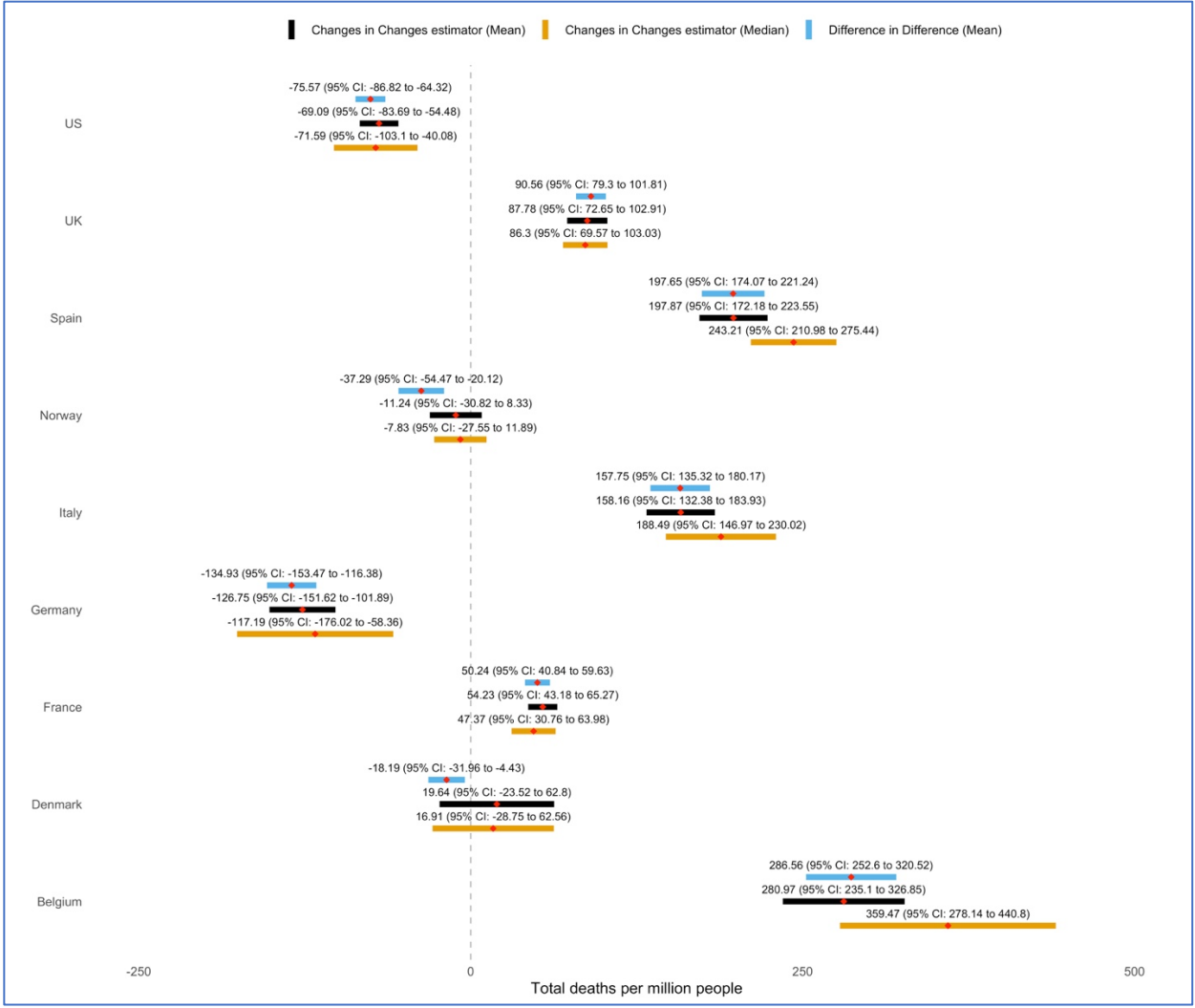

Notes: The standard errors are computed using bootstrap estimation with 50 replications. 95% Confidence intervals are in parenthesis.

**Figure 4:** Changes-in-Changes (CIC) and Differences-in-Differences (DID) estimates for daily deaths per million people, for each country with Sweden as the control group

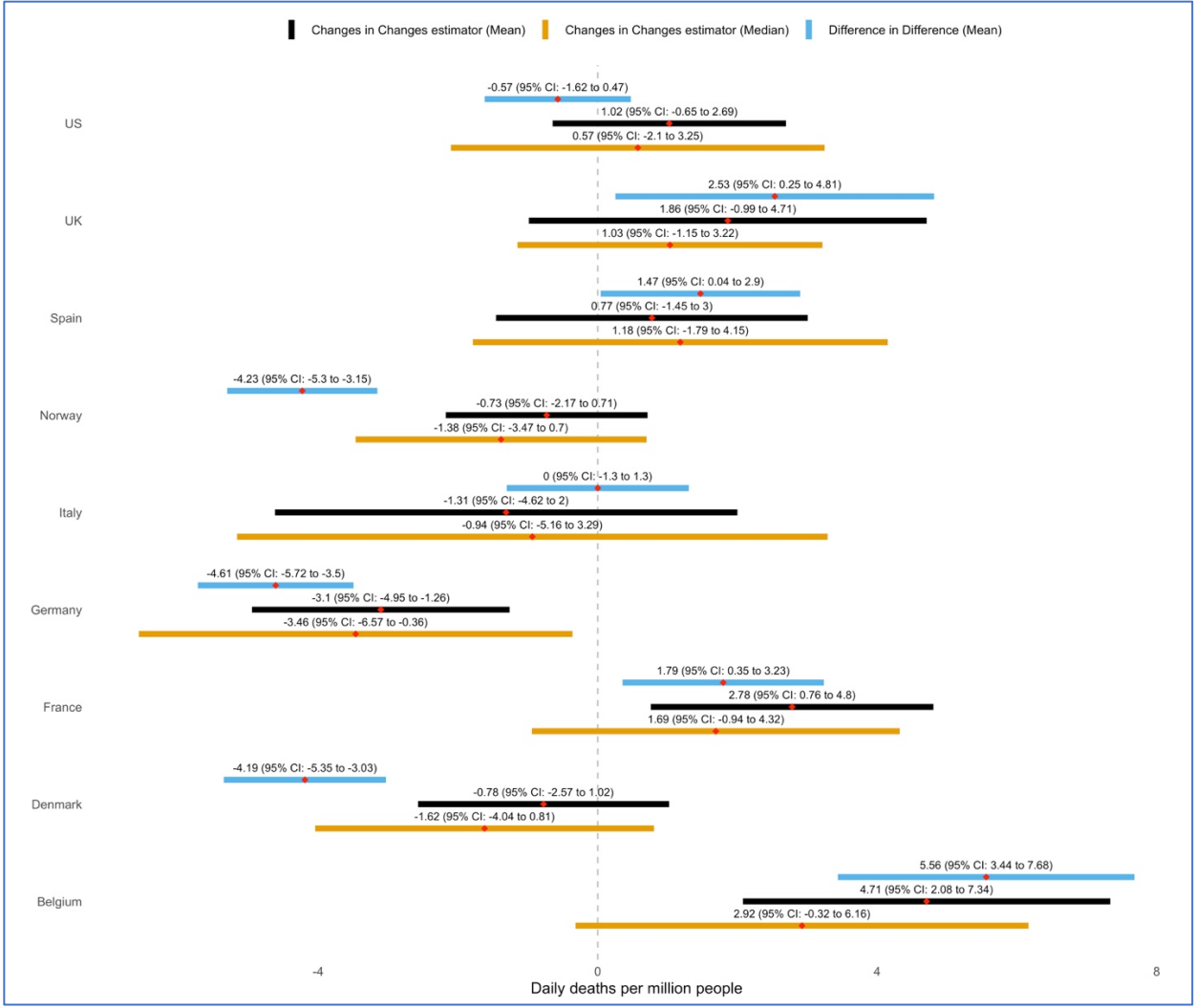

Notes: The standard errors are computed using bootstrap estimation with 50 replications. 95% Confidence intervals are in parenthesis.